\documentclass[
reprint,
 aps,
 superscriptaddress,
]{revtex4-2}

\usepackage{physics}
\usepackage{dsfont}
\usepackage{color,newfloat}
\usepackage{graphicx}
\usepackage{dcolumn}
\usepackage{bm}
\usepackage{amsmath,amssymb,amsthm}
\usepackage{mathtools}
\usepackage{multirow}
\usepackage[dvipsnames]{xcolor}
\usepackage{hyperref}
\hypersetup{
    colorlinks=true,
    linkcolor=Maroon,
    citecolor=Maroon,
    urlcolor=Maroon
}
\usepackage{siunitx}

\begin{document}


\title{
High-speed thin-film lithium niobate quantum processor\\ driven by a solid-state quantum emitter
}

\author{Patrik I. Sund}
\altaffiliation{Equally contributing authors.}
\affiliation{Center for Hybrid Quantum Networks (Hy-Q), Niels Bohr Institute, University of Copenhagen, Blegdamsvej 17, DK-2100 Copenhagen, Denmark}

\author{Emma Lomonte}
\altaffiliation{Equally contributing authors.}
\affiliation{Institute of Physics, University of Muenster, 48149, Muenster, Germany}
\affiliation{CeNTech — Center for Nanotechnology, 48149, Muenster, Germany}
\affiliation{SoN — Center for Soft Nanoscience, 48149, Muenster, Germany}

\author{Stefano Paesani}
\email{stefano.paesani@nbi.ku.dk}
\affiliation{Center for Hybrid Quantum Networks (Hy-Q), Niels Bohr Institute, University of Copenhagen, Blegdamsvej 17, DK-2100 Copenhagen, Denmark}

\author{Ying Wang}
\affiliation{Center for Hybrid Quantum Networks (Hy-Q), Niels Bohr Institute, University of Copenhagen, Blegdamsvej 17, DK-2100 Copenhagen, Denmark}

\author{Jacques Carolan}
\affiliation{Center for Hybrid Quantum Networks (Hy-Q), Niels Bohr Institute, University of Copenhagen, Blegdamsvej 17, DK-2100 Copenhagen, Denmark}
\affiliation{Wolfson Institute for Biomedical Research, University College London, London, UK}

\author{Nikolai Bart}
\author{Andreas D. Wieck}
\author{Arne Ludwig}
\affiliation{Lehrstuhl f{\"u}r Angewandte Festk{\"o}rperphysik, Ruhr-Universit{\"a}t Bochum, Universit{\"a}tsstrasse 150, D-44780 Bochum, Germany}

\author{Leonardo Midolo}
\affiliation{Center for Hybrid Quantum Networks (Hy-Q), Niels Bohr Institute, University of Copenhagen, Blegdamsvej 17, DK-2100 Copenhagen, Denmark}

\author{Wolfram H. P. Pernice}
\affiliation{Institute of Physics, University of Muenster, 48149, Muenster, Germany}
\affiliation{CeNTech — Center for Nanotechnology, 48149, Muenster, Germany}
\affiliation{SoN — Center for Soft Nanoscience, 48149, Muenster, Germany}
\affiliation{Heidelberg University, Im Neuenheimer Feld 227, 69120 Heidelberg, Germany}

\author{Peter Lodahl}
\affiliation{Center for Hybrid Quantum Networks (Hy-Q), Niels Bohr Institute, University of Copenhagen, Blegdamsvej 17, DK-2100 Copenhagen, Denmark}

\author{Francesco Lenzini}
\email{lenzini@uni-muenster.de}
\affiliation{Institute of Physics, University of Muenster, 48149, Muenster, Germany}
\affiliation{CeNTech — Center for Nanotechnology, 48149, Muenster, Germany}
\affiliation{SoN — Center for Soft Nanoscience, 48149, Muenster, Germany}


\begin{abstract}

Scalable photonic quantum computing architectures pose stringent requirements on photonic processing devices. 
The need for low-loss high-speed reconfigurable circuits and near-deterministic resource state generators are some of the most challenging requirements.
Here we develop an integrated photonic platform based on thin-film lithium niobate and interface it with deterministic solid-state single-photon sources based on quantum dots in nanophotonic waveguides. 
The generated photons are processed with low-loss circuits programmable at speeds of several GHz.
We realize a variety of key photonic quantum information processing functionalities with the high-speed circuits, including on-chip quantum interference, photon demultiplexing, and reprogrammability of a four-mode universal photonic circuit. 
These results show a promising path forward for scalable photonic quantum technologies by merging  integrated photonics with solid-state deterministic photon sources in a heterogeneous approach to scaling up.
\end{abstract}

\date{\today}

\maketitle



Sustained progress of quantum technologies has enabled quantum hardware to compete with and surpass the capabilities of classical supercomputers~\cite{arute2019, wu2021, zhong2020quantum, madsen2022quantum}.
However, significant challenges remain in order to reach controllable quantum systems at scale, as required for currently known practical applications and fault-tolerant quantum technologies~\cite{preskill2018}.
Photonics is a promising platform for unlocking scalable quantum hardware~\cite{wang2020, uppu2021quantum},
with the capacity to enable long-range quantum networks~\cite{yin2017, pompili2021}, interconnections between multiple quantum devices~\cite{llewellyn2020}, and large-scale photonic circuits for quantum computing and simulation~\cite{harris2017, wang2018multidimensional, paesani2019, arrazola2021}.
Central to photonic quantum technologies is the generation of high-quality photonic states and fast low-loss programmable circuits for their routing and processing~\cite{rudolph2017, bombin2021}. 
In recent years, significant progress has been achieved in developing near-ideal high-efficiency sources of indistinguishable photons, with solid state quantum emitters, such as quantum dots (QDs), standing out as excellent candidates for the realization of on-demand single-photon sources (SPSs)~\cite{Ding2016, somaschi2016, uppu2020scalable, tomm2021}. 
Low-loss optical circuits that can be rapidly reprogrammed (typically GHz rates required) are central for key tasks in photonic quantum information processing, such as multiplexing or demultiplexing of single-photon sources~\cite{xiong2016, lenzini2017, kaneda2019}, routing for fusion networks and feed-forward operations for fusion-based quantum computation~\cite{bombin2021, bartolucci2021fusion}. 
However, a platform satisfying all requirements for the integration of GHz-speed low-loss circuits at scale has so far been elusive.
Indeed, standard approaches, such as thermal modulators or MEMS in the SOI or SiN platforms, are limited to operating speeds up to MHz~\cite{harris2014, Errando2019, arrazola2021}, while high-speed free-carrier modulators add significant phase-dependent photon loss~\cite{BaehrJones12}.
Very recently, programmable integrated circuits based on piezo-optomechanical actuators coupled to SiN waveguides with a modulation speed up to approximately 120 MHz were demonstrated both at room and cryogenic temperatures~\cite{dong2022}. Still, the scalability of these devices is currently limited by high insertion loss ($> 3 \ \mbox{dB}$) and a low modulation efficiency (voltage-length product $\simeq 50 \ \text{V} \cdot \text{cm}$).
In this context, single crystal thin lithium niobate (LN, LiNbO$_3$) films bonded on a silica insulating substrate (LNOI: Lithium-Niobate-On-Insulator) have emerged in recent years as a particularly promising platform.
Thanks to the strong electro-optical properties, high transparency, and high index contrast, integrated circuits with compact footprints and low propagation loss can be implemented with tunable phase-shifters having a modulation bandwidth above several GHz~\cite{zhang2017,wang2018Nanophotonic,wang2018-2}. 
The wide transparency range of LN makes these circuits also especially attractive for operation with a variety of solid-state quantum emitters, which are characterized by an emission spectrum typically at visible or near-infrared wavelengths~\cite{aharonovich2016solid}. Moreover, electro-optic modulators (EOMs) implemented in LN are compatible with operation at cryogenic temperature~\cite{herzog2008,thiele2020,lomonte2021}, making full-system integration with QDs and high-quality superconducting nanowire single-photon detectors (SNSPDs) an additional promising prospect~\cite{aghaeimeibodi2018, lomonte2021, Christen2022}. 
However, programmable multimode LNOI circuits for quantum information processing at the single photon regime have yet to be demonstrated.
Here we report on the experimental realization of a reconfigurable LNOI quantum photonic processor for the control and operation of quantum states of light emitted from a QD single-photon source.
The device is designed to feature low propagation loss, efficient coupling to a QD photon emitter via a fiber-to-chip interface in the targeted 900-950 nm wavelength range (typical emission wavelength of InGaAs QDs), and modulators that enable programming the quantum processor up to GHz speed. 
By injecting single photons emitted by a waveguide-integrated QD source into the LNOI optical circuitry, we show prototypical key functionalities needed for photonic quantum information processing, including multiphoton interference, active four-spatial-mode routing and demultiplexing of a temporal stream of photons, and the operation of a 4$\times$4 reconfigurable universal unitary circuit~\cite{clements2016}.


\section{Results}

\subsection{Integrated photonic platform}

\begin{figure*}[thpb]
    \centering
    \includegraphics[width=\textwidth]{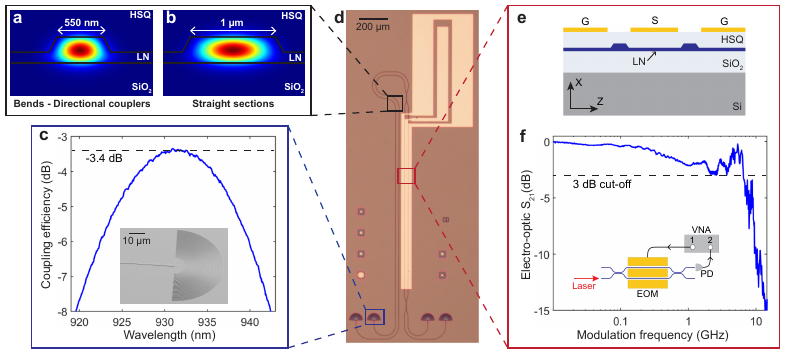}
    \caption{
    \textbf{Overview of the platform.}
    \textbf{a-b.} Schematics of the designed waveguide geometry, tailored for the quantum emitter $\lambda\simeq940$~nm operation wavelengths, for \textbf{(a)} single-mode waveguides used in bends and directional couplers, and \textbf{(b)} multi-mode straight waveguides. 
    Color-coded is the field intensity of the fundamental TE waveguide mode. 
    \textbf{c.} Measured coupling efficiency of the fabricated grating couplers as a function of the input laser wavelength, with a peak efficiency of -3.4 dB. The inset shows a scanning electron micrograph image of the coupler. 
    \textbf{d.} Optical microscope image of an electrically tunable MZI.
    \textbf{e.} Schematic of the cross-section of the electro-optic phase shifter.
    \textbf{f.} Modulation bandwidth of the MZI measured with a VNA. 
    The data shows a 3 dB cut-off at approximately 6.5 GHz.
    Inset: schematic of the setup employed in the measurement.}
    \label{fig:platform}
\end{figure*}
In Fig.~\href{fig:platform}{\ref{fig:platform}a} we report a schematic of the geometry employed for the realization of single-mode (SM) LNOI waveguides. 
Optical circuits are implemented as rib waveguides with a 180 nm etching depth, fabricated by electron-beam lithography (EBL) and argon etching on a 300 nm thick X-cut LN film bonded on a silica-on-silicon substrate (see Methods for more details). 
After etching, the waveguides are clad with a $\simeq 550$ nm thick electrically cured hydrogen silsesquioxane (HSQ) layer.
The angle of the waveguide sidewalls (measured with an AFM) is $\simeq 60^\circ$, and the top waveguide width of 550 nm ensures single-mode operation around 940 nm wavelength, the typical emission wavelength for InGaAs QDs employed in this work, for TE-polarized light.
The fabrication of photonic integrated circuits (PICs) operating at the short emission wavelength of InGaAs QDs comes with the disadvantage of increased propagation loss due to sidewall scattering, whose magnitude scales inversely with the third power of the operation wavelength \cite{lacey1990}, when compared with the values attainable for telecom LNOI waveguides. For this reason, we chose to employ SM waveguides only for the realization of bends and directional couplers---where it is more likely to excite higher order modes---and to adiabatically enlarge the waveguide width up to 1 $\text{\textmu}$m in the straight sections of the optical circuitry to reduce the overlap between the TE optical mode and the waveguide sidewalls (see Fig.~\href{fig:platform}{\ref{fig:platform}b}). By the use of this approach, we were able to measure a low propagation loss coefficient down to 0.84 dB/cm at $\lambda=940 \ \mbox{nm}$ (see Supplementary Information for more details).
The LNOI PIC is optically coupled to SM fibers by the use of apodized grating couplers with a negative diffraction angle, designed by following the approach of Ref.~\cite{lomonte2021-2}. 
The best coupling efficiency of our gratings was measured around 930 nm wavelength, and found to be -3.4 dB (see Fig.~\href{fig:platform}{\ref{fig:platform}c} and Methods for further details). This measured value compares well with the best ones obtained at telecom wavelength for gratings etched on a pure LNOI platform, i.e., without any use of a back-reflector for improving the grating directivity or additional material layers for increasing the grating strength (see, e.g., Ref.~\cite{chen2022} for a review on recent progress). 
We numerically estimate that with the aid of a metal back-reflector underneath the buried oxide layer, our couplers can achieve an insertion loss of $< 1 $ dB. 
The LNOI PICs demonstrated here can therefore offer an  efficient approach  for interfacing fast optical switches and circuits to optical fibers.
For the realization of electro-optically tunable waveguide circuits we use as a main building block Mach--Zehnder interferometers (MZIs) consisting of two 50:50 directional couplers and an electrically tunable phase shifter (see Fig.~\href{fig:platform}{\ref{fig:platform}d}). 
The phase shifter is implemented by patterning three 1.25 mm long gold electrodes in a ground--signal--ground configuration along the Y axis of the crystal, in order to provide an efficient overlap between the fundamental TE mode of the waveguide and the Z-component of the applied electric field via the highest electro-optic component ($r_{33}\simeq 30 \ \text{pm/V}$) of the LN susceptibility tensor. 
Unlike more common implementations of EOMs in LNOI, where signal and ground electrodes sit at the two sides of the waveguide \cite{wang2018Nanophotonic,wang2018-2}, here we opted to pattern them atop of the HSQ cladding (see Fig.~\href{fig:platform}{\ref{fig:platform}e}). 
This choice is made to enable a nearly lossless crossing of the electrodes with the waveguides without any need for additional fabrication steps. 
The fabricated modulators displayed a half-wave voltage ($V_{\pi}$) of approximately $ \SI{4.5}{\volt} $, corresponding to a small voltage-length product $V_{\pi}L \simeq 0.6 \ \text{V} \cdot \text{cm}$, and a high extinction ratio $\simeq 21 \ \text{dB}$.
The insertion loss of a single MZI, which includes two directional couplers, propagation loss in the waveguides, and metal-induced absorption loss due to the presence of the electrodes, was estimated equal to $\simeq$0.8 dB (see Methods).
To test the high-speed performance of the modulators, we make use of the setup schematically depicted in the inset of Fig.~\href{fig:platform}{\ref{fig:platform}e}. 
The phase shifter is driven with a small-amplitude RF signal from port 1 of a vector network analyzer (VNA), while port 2 is connected to a fast photodiode (Newport - 1544 B) optically coupled to one of the two outputs of the MZI.
The resulting $S_{21}$ parameter---defined as the ratio between the power measured at port 2, and the power leaving from port 1---is plotted in Fig.~\href{fig:platform}{\ref{fig:platform}f} normalized to its maximum value, and provides a direct estimation of the MZI electro-optic bandwidth. From the data, we record a 3 dB cut-off frequency of the modulator at approximately $6.5 \ \text{GHz}$, demonstrating the high-speed capabilities of the fabricated PICs.

\subsection{On-chip quantum interference}

\begin{figure}
    \includegraphics[width=0.48\textwidth]{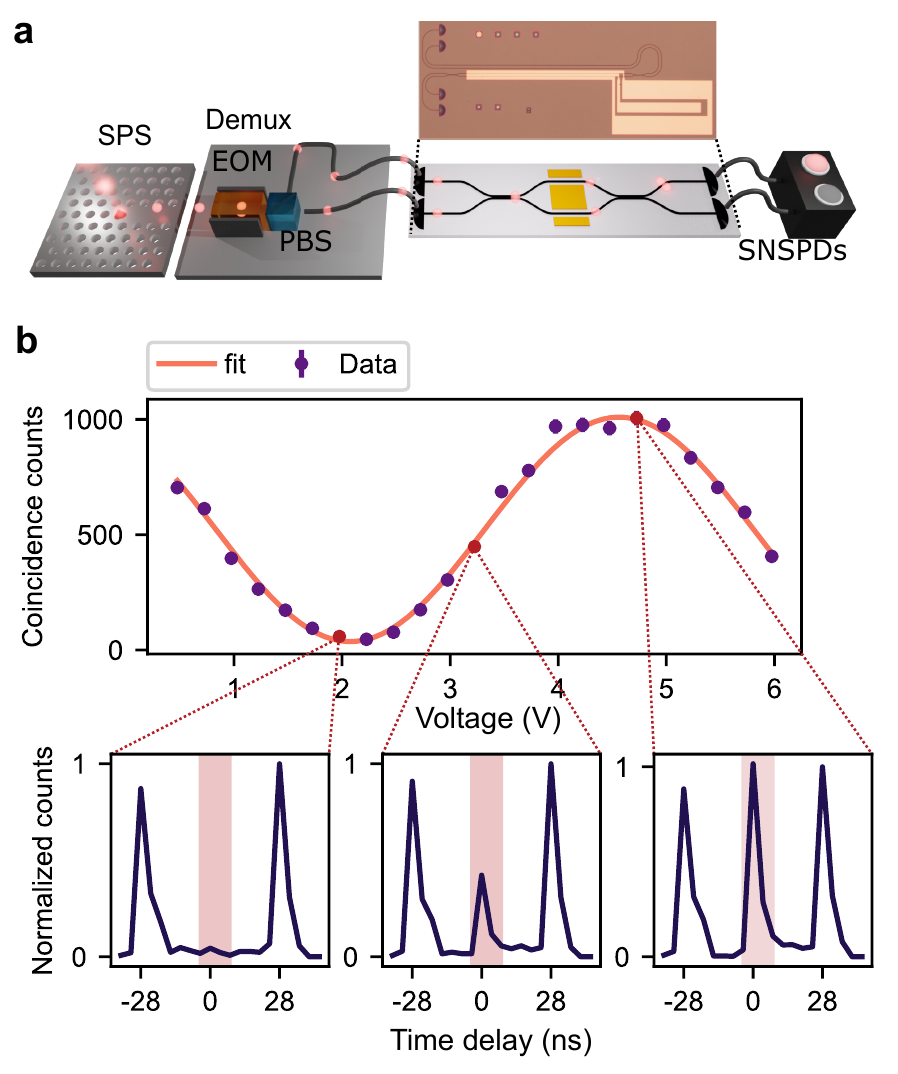}
    \caption{
    \textbf{Measurement of on-chip quantum interference.}
    \textbf{a.} Schematic of the experimental setup. Photons generated by a QD single-photon source (SPS) are sent into a two-mode demultiplexer consisting of a resonantly enhanced electro-optic modulator (EOM) and a polarizing beam-splitter (PBS).
    The photons are subsequently collected into fibers and injected into the LNOI chip by a fiber array.
    Controlling the delay on one of the demultiplexer arms ensures that the photon pairs arrive at the device simultaneously, and fiber polarization controllers are used to optimize coupling into the TE mode.
    The output photons are collected via the same fiber array and routed to SNSPDs for coincidence detection.
    Inset: image of the MZI device used.
    \textbf{b.} Recorded coincidence data at zero time-delay (shaded red areas in the insets) for varying applied voltages. 
    Minima and maxima in the observed HOM fringe correspond to applied phases of $\phi_\text{min} = \pi/2 + k \pi$ and $\phi_\text{max} = k \pi$, respectively, with $k$ an integer number.
    The error bars are estimated from Poissonian statistics and are smaller than the data points.
    The HOM visibility of the quantum interference is determined from a curve fit (orange line) to be $92.7\% \pm 0.7\%$.
    Insets: coincidence histograms for three different applied voltages.
    }
    \label{fig:HOM}
\end{figure}

A central figure of merit in photonic quantum information processing is the visibility of multiphoton quantum interference, typically quantified as the two-photon interference visibility in Hong--Ou--Mandel (HOM) experiments.
Incorporating simultaneously the detrimental effects of imperfect photonic circuitry, source distinguishability, and purity, this quantity is essential in determining the dominant stochastic noise in photonic quantum computing schemes~\cite{rudolph2017} and the computational complexity limits in photonic sampling algorithms~\cite{Renema2018}.
We thus start by performing on-chip HOM experiments to test the performance of our platform for photonic quantum information processing with the experimental scheme depicted in Fig.~\href{fig:HOM}{\ref{fig:HOM}a}.
We create a two-photon input state from a stream of single photons emitted by the QD using an off-chip demultiplexer to split up pairs of consecutive photons into two distinct paths, one of which is delayed such that the photons arrive at the chip simultaneously. 
The photons are sent into an integrated MZI containing a tunable electro-optic phase-shifter. 
The electrodes constituting the phase shifter are connected to an electronic probe, allowing for control of the internal phase in the MZI. 
Photons are then routed off-chip to SNSPDs for coincidence detection. 
On-chip HOM interference is studied by applying a varying phase to the high-speed modulator of the MZI interferometer, and measuring the fringe of the coincidence counts at the output~\cite{rarity1990two, adcock2019, paesani2020}.
The measured HOM fringe visibility of $92.7\% \pm 0.7\%$, see Fig.~\href{fig:HOM}{\ref{fig:HOM}b}, is consistent with the corresponding HOM visibility measured off-chip (see Methods). This testifies that the fabricated PIC does not add stochastic noise on the processed photonic states (e.g., due to imperfect phase shifting or beam splitting, temporal mismatches, excitation of higher order modes in the waveguides, or TE-TM intermodal conversion~\cite{kaushalram2020mode}), certifying the high quality of the developed circuits as quantum photonic processing units. 

\subsection{Integrated single-photon router}

\begin{figure*}
    \includegraphics[width=\textwidth]{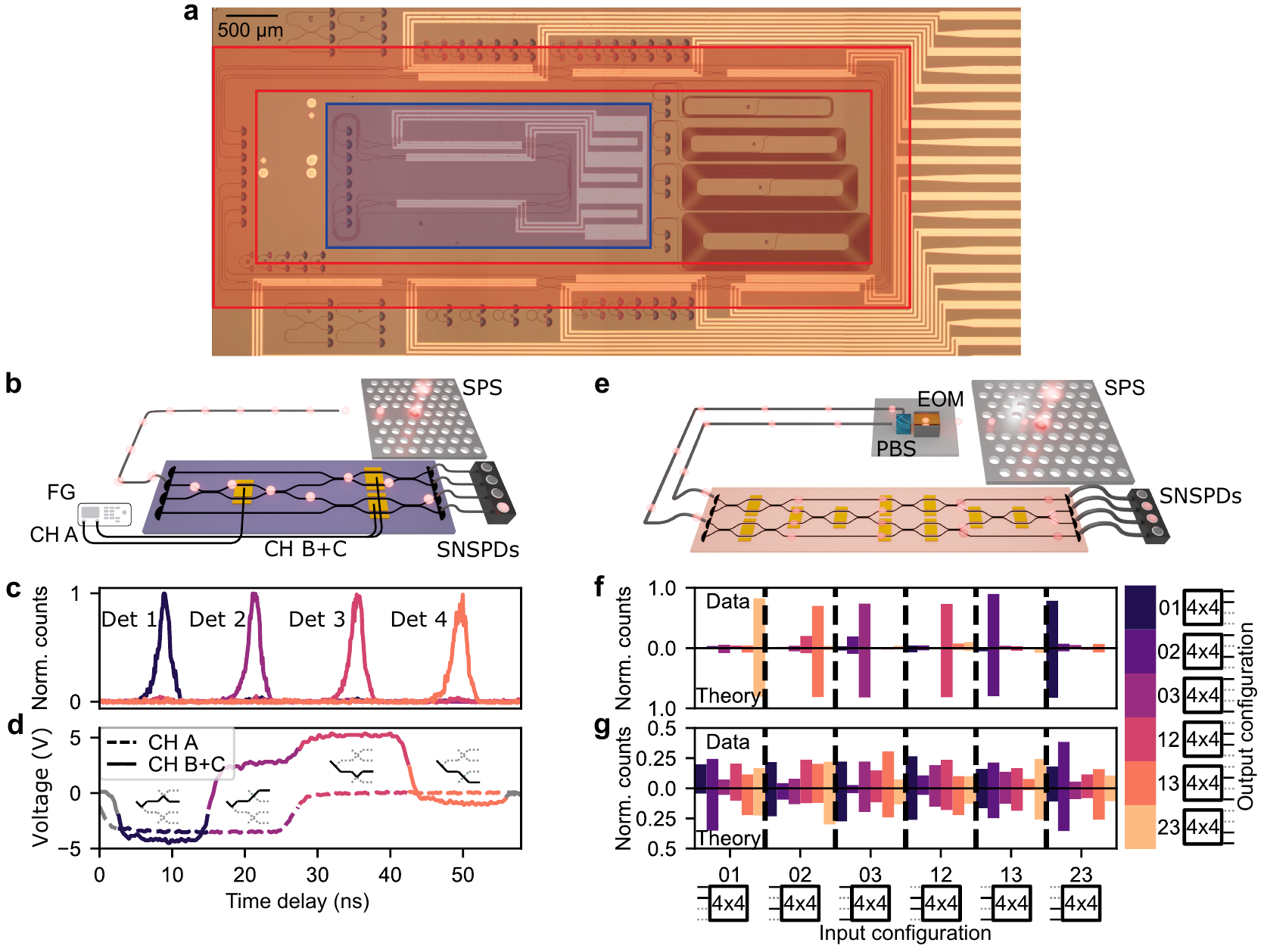}
    \caption{
    \textbf{Photon processing in multimode high-speed integrated circuits.}
    \textbf{a.} Optical image of the chip. 
    The photon router structure is highlighted in blue, and the $4\times 4$ universal interferometer in orange. 
    \textbf{b.} Schematic of the experimental setup used to perform active $1\times 4$ demultiplexing of a stream of single-photons produced by the QD. 
    Photons are directly coupled in and out of the chip using a fiber array, and their time of arrival recorded via SNSDPs and a time-tagger. 
    Fast electrical control is performed via a function generator (FG) connected to the modulators via a probe station, where a channel is used to individually address the MZI in the first layer and the other channel is split to drive both MZIs in the second layer in parallel.
    \textbf{c.} Normalized photon counts in the four output waveguides within the time interval of a four-photon sequence. 
    \textbf{d.} Associated pulse sequences, with corresponding switching network configurations shown as insets.
    \textbf{e.} Schematic of the experimental setup for the universal $4 \times 4$ interferometer.
    The optical part is equivalent to that described in Fig.~\href{fig:HOM}{\ref{fig:HOM}}.
    The ten high-speed modulators employed are electrically connected via a probe station and driven by a multichannel function generator.
    \textbf{f-g.} Experimental data (top) and estimated theoretical (bottom) collision-free input--output probability distributions when programming the interferometer to implement an approximate permutation matrix (\textbf{f}) and a randomized unitary matrix (\textbf{g}), with estimated statistical fidelities of $96.3\%$ and $95.5\%$, respectively. 
    The horizontal index indicates the input configuration in terms of the mode indices of the first and second photon, and the color corresponds to the output configuration.
    Dashed lines separate different input configurations.
    }
    \label{fig:demux+4x4}
\end{figure*}

Fast photon routers play an important role in photonic quantum computing schemes. 
For example, routers can be used in conjunction with measurement and feed-forward to construct multiplexing schemes, turning inherently probabilistic quantum photonic processes into near-deterministic operations~\cite{Migdall2002, Varnava2008, GimenoSegovia2015}. 
Alternatively, taking advantage of the capability of deterministic quantum emitters, streams of emitted photons can be routed into multiple spatial outputs to enable networking schemes that significantly reduce resource overheads in photonic quantum computing architectures~\cite{bombin2021}, as well as demultiplexing schemes that turn a single deterministic single-photon source into multiple sources~\cite{Wang2019Boson}.

Making use of the capability to integrate fast phase shifters on the LNOI platform compatible with the quantum emitter wavelengths, we demonstrate a fully on-chip photon router for the QD-emitted photons. 
In particular, starting from a stream of single photons emitted from the QD at a fixed rate, we implement a $1\times 4$ demultiplexer.
The demultiplexer consists of three fast electro-optic MZI switches, which are cascaded in a tree-shaped network, as schematised in  Fig.~\href{fig:demux+4x4}{\ref{fig:demux+4x4}b} (see Fig.~\href{fig:demux+4x4}{\ref{fig:demux+4x4}a} for a device microscope image).
It processes sequences of four single photons emitted by the QD with a temporal separation of $\SI{13.8}{\nano\second}$, switching each photon deterministically into its own dedicated spatial mode.
Fig.~\href{fig:demux+4x4}{\ref{fig:demux+4x4}c} shows time traces for four output detectors with respect to the pulse sequence. 
The average probability of switching a photon in the four-photon sequence 
to its dedicated mode is measured to be $96.2\%$, corresponding to an average suppression of $-14.2$~dB for unwanted photons. 
This directly illustrates the highly promising potential of the LNOI platform for photon routing of deterministic resource states produced by QDs.

\subsection{Universal four-mode interferometer}

Programmable multimode quantum photonic interferometers are central in the implementation of core functionalities in photonic quantum technologies, e.g., multi-photon gates and fusion measurements~\cite{carolan2015, bartolucci2021fusion}, and to realize circuits for quantum computational advantage experiments or analog quantum simulation~\cite{zhong2020quantum, madsen2022quantum, Huh2014, Sparrow2018}.
To showcase the capabilities of the QD-LNOI platform for this class of tasks, we implement a universal $4\times 4$ interferometer constructed from a network of 6 MZI and 10 phase modulators, as shown in Fig.~\href{fig:demux+4x4}{\ref{fig:demux+4x4}e}. 
The interferometer implements a scheme from Clements et al.~\cite{clements2016}, whereby the unitary transformation enacted by the interferometer is controlled by the phases applied to the modulators, allowing for the device to implement any arbitrary linear-optical unitary transformation on the four input waveguides.

We show that the interferometer can be programmed from implementing a structured matrix (an approximate permutation matrix resulting from applying no voltage to the modulators) to implementing a randomized matrix obtained by driving all the phase shifters simultaneously (see Supplementary Information).
In Figures~\href{fig:demux+4x4}{\ref{fig:demux+4x4}f, g} we report the measured input--output probability distributions for the two cases, obtained by photon coincidence detection for all possible collision-free two-photon configurations, i.e. combinations with no more than one photon per mode. 
The different input states were prepared by demultiplexing a stream of photons from the QD off-chip into two separate fibers and routing these two fibers to all possible collison-free two-photon input configurations (see Fig.~\href{fig:demux+4x4}{\ref{fig:demux+4x4}e}).
The measured distributions are compared with theoretical predictions obtained by reconstructing the unitary matrices from the experimental data (see Supplementary Information for more details).
We find a mean statistical fidelity to the implemented transformation of $96.3\%$ for the \mbox{approximate} permutation matrix, and of $95.5\%$ for the \mbox{randomized matrix}. 

\section{Discussion}

The demonstrated exemplary circuits show the promise of the developed LNOI platform for processing photons from emerging solid-state deterministic sources. 
In order to fully enable scalable quantum technologies, further optimization of the platform is, however, required.
To improve the transmission loss at these wavelengths, besides improving our current fabrication process to reduce roughness, the use of a thinner LN film would enable the realization of SM waveguides with a larger width and to mitigate the effect of sidewall \mbox{scattering}. 
The use of a cladding with a higher refractive index than HSQ (e.g., silicon oxynitride) can also be beneficial, at the expense of a larger waveguide footprint due to the lower index contrast. 
A large improvement in the efficiency of the grating couplers can be achieved by patterning a metal back-reflector underneath the buried oxide layer. 
This can, for example, be accomplished via back-side etching of the silicon handle, with a fabrication process analogous to the one preliminary demonstrated in Ref.~\cite{lomonte2022-3}. 
System integration of all the employed quantum devices on LNOI ---sources, circuits, and detectors---provides an additional promising direction for lowering the overall system loss. 
In this direction, direct integration of SNSPDs together with a reconfigurable LNOI component has been recently experimentally demonstrated using the same material stack employed in this work~\cite{lomonte2021}. 
Heterogeneous integration of quantum emitters directly onto the LNOI waveguides could be realized with pick-and-place techniques that has recently emerged as a highly promising approach~\cite{aghaeimeibodi2018}. 
Alternatively, photonic wire bonds fabricated by two-photon absorption lithography might be employed for interfacing optical circuits implemented in heterogeneous photonic platforms~\cite{lindenmann2012}. 
In conclusion, high-speed LNOI quantum processors provide a route to scale up quantum photonic technologies by leveraging the mature and advanced on-demand sources realized with solid-state quantum emitters in photonic nanostructures. 
Moving forward, further optimization of the platform is required in order to reduce coupling and propagation loss, which will be realized in a close interplay between device design and PIC fabrication optimization. 
Fault-tolerant quantum computing architectures demand typical loss levels of $\lesssim 10\%$ per photon~\cite{Borregaard2020, bartolucci2021fusion}, which appears feasible with the technology after full-circle optimization of source, processor, and detector, and will be an exciting future research and engineering challenge.
With such an approach, only a few tens of interconnections between individual quantum emitters and GHz-speed reconfigurable devices suffice for fault-tolerant photonic quantum computing at scale~\cite{bombin2021}.
%


\bibliography{biblio.bib}

\ \\
\noindent\textbf{Acknowledgements.} 
We thank Ravitej Uppu, Kasper H. Nielsen, and Carlos Faurby for valuable discussions and technical assistance. We would like to thank the Muenster Nanofabrication Facility (MNF) for their support in nanofabrication-related matters.
We acknowledge funding from the Danish National Research Foundation (Center of Excellence “Hy-Q,” grant number DNRF139) and the Novo Nordisk Foundation (Challenge project "Solid-Q").
F.L., E.L., and W.P. acknowledge funding from the European Union's Horizon project CLUSTEC, and the BMBF projects PhoQuant and MUNIQC-Atoms. 
S.P. acknowledges funding from the Cisco University Research Program Fund (nr. 2021-234494) and from the Marie Skłodowska-Curie Fellowship project QSun (nr. 101063763).
N.B., A.D.W., and A.L. acknowledge funding through the BMBF QR.X Project 16KISQ009.


\section{Methods}

\subsection{Fabrication of the LNOI PICs}
The LNOI PICs are fabricated on a 300 nm thick X-cut LN film bonded on a silica-on-silicon wafer, with a Si thickness of 500 $\mbox{\textmu m}$, and a SiO$_2$ thickness of 4.7 $\mbox{\textmu m}$ (wafer produced by NanoLN). As a first step, gold markers for EBL alignment are patterned on the plain LN film with a standard lift-off process. Photonic circuits are afterwards patterned by EBL with a negative resist (AR-N 7520.18), followed by Ar etching with an ICP system. After etching and resist stripping, a RCA-1 cleaning process is employed to remove the redeposition of sputtered material on the waveguide sidewalls caused by the physical Ar etching. Next, the waveguides are clad with a $\simeq 550$ nm thick electrically cured HSQ layer, and annealed at $400 \ ^{\circ} \mbox{C}$ for one hour to reduce the absorption loss of the LN crystal. As a last step, the modulators are patterned atop of the HSQ cladding with a standard lift-off process.  

\subsection{Solid-state quantum emitter details}
Self-assembled indium arsenide (InAs) quantum dots (QDs) are adopted to generate single photons. A GaAs pin diode structure was fabricated on a sacrificial $\text{Al}_{.75}\text{Ga}_{.25}\text{As}$ layer, resulting in a 180 nm thick membrane. QDs were introduced in the middle of the membrane. Photonic and electronic nanostructures to interface with the QDs are fabricated via mature fabrication techniques developed for this solid-state platform~\cite{uppu2021quantum}. Firstly, high-quality metal gates are fabricated and used to suppress electronic noise, as well as for tuning the QDs' emission wavelength via Stark-shift ~\cite{uppu2020scalable,Freja2020}. To improve the single photon collection efficiency, single-sided photonic crystal waveguides were employed ~\cite{Ying2022}, which were terminated with a row of mirroring holes reflecting the single photons emitted on the reflection side of the waveguide. On the other side of the waveguide a shallow-etched grating coupler is used for off-chip coupling, with a distributed Bragg reflector (DBR) grown below the sacrificial layer that efficiently shapes the mode for coupling to a single-mode fiber. With this source, we generate 15.6 million photons/second in a fiber for a QD source pumped at a repetition rate at 72.6 MHz ~\cite{Ying2022}, where residual photon losses are dominated by optical elements in the collection path and can be readily improved even further.
The sample was placed in a 1.6 K cryostat and pulsed resonant excitation was used to achieve high-quality single-photon generation. The source was characterized with a standard free space single-photon spectroscopy setup to achieve a near-unity single-photon purity of $g^{(2)}(0)=0.005\pm0.001$, i.e. 99.5$\%$ single-photon purity, and a degree of indistinguishability of $94.5\%\pm1.7\%$ between subsequently emitted photons in a Hong--Ou--Mandel interference experiment. 

\subsection{Measurement of the grating couplers efficiency}
The efficiency per coupler reported in Fig.~\href{fig:platform}{\ref{fig:platform}c} is estimated as the square root of the transmission of a device consisting of only two grating couplers connected by a short waveguide (length of $\simeq 300 \ \mbox{\textmu m}$). Optical coupling to the gratings is achieved by the use of an angled polished fiber-v-groove array equipped with SM fibers for the 900-950 nm wavelength range (fiber type: 780HP). The insertion loss of the fiber array (including fiber connector loss and Fresnel loss at the output facet of the array) is calibrated out from the transmission measurement.  

\subsection{Measurement of the modulators insertion loss}
The insertion loss of an electrically tunable MZI is estimated by measuring the transmission of the demultiplexer of Fig.~\href{fig:demux+4x4}{\ref{fig:demux+4x4}b} for all waveguide inputs. When light is injected into one of the two external inputs, it goes through only one MZI. When light is injected into one of the two internal inputs, it goes through two MZIs. Comparing the total transmission over all outputs between these two cases we find an average MZI insertion loss equal to 0.8 dB $\pm$ 0.3 dB. This measurement via the demultiplexer structure provides an estimate that is more resilient to variations in the efficiency of other components (e.g. grating couplers and loss in routing waveguides) compared to measuring a single MZI structure (which provided a compatible measurement). 
We numerically estimate that the insertion loss of our modulators can be lowered to approximately 0.2 dB, at the expense of an increase of $\simeq 2$ V in the half-wave voltage, by carefully adjusting the thickness of the cladding and the gap between the electrodes in order to reduce the amount of metal-induced propagation loss. A further improvement below this value will require an optimization of our fabrication workflow for reducing the effect of sidewall scattering (see the Discussion section).

\end{document}